 \documentclass[rnote]{aa} 
\usepackage{graphicx}
\usepackage{txfonts}
\begin{document}
   \title{SDSS photometry of Asteroids in Cometary Orbits}


   \author{A. Alvarez-Candal
          \inst{1,2}
          }

   \institute{ESO, Alonso de C\'ordova 3107, Vitacura, Casilla 19001, Santiago 19, Chile.\\
              \email{aalvarez@eso.org}
         \and
             Insituto de Astrof\'\i sica de Andaluc\'ia - CSIC, Glorieta de la Astronom\'ia s/n, E18008, Granada, 
Spain.
             }

   \date{Received ---; accepted ---}

 
  \abstract
   {}
{{{\bf Aiming at exploring the relationship between asteroids in cometary orbits and other minor body
populations from the observational point of view, I explore the large photometric database of the 
Sloan Digital Sky Survey - Moving Objects Catalog.}}
}
   {The sample of interest was carefully selected analysing colors and orbital 
properties within the data in the catalog. I computed the spectral slope for each object 
and compared with published spectroscopic results of other Asteroids in Cometary Orbits, 
as well as with other populations of asteroids in the outer region of the Main belt and Trojans.}
   {Using this extended database I find that Asteroids in Cometary Orbits with 
{{\bf Tisserand parameter below 2.9, and especially below 2.8, are most likely of primitive origin.
}}
The population of objects with Tisserand parameter larger than 2.9 is a mixture of populations
ranging from the inner to the outer Main belt.}
   {}


   \maketitle

\section{Introduction}

An ``Asteroid in Cometary Orbit'', ACO, is an asteroid-like (no apparent coma) 
object in a cometary-like orbit.
Classically a cometary orbit has $T_{\rm J}<3$, where $T_{\rm J}$ is the {\bf Tisserand parameter
with respect to Jupiter} and is given by  
\begin{equation}\label{eq01}
T_{\rm J} = a_{\rm J}/a + 2 \cos{I}\sqrt{(a/a_{\rm J})(1-e^2)},
\end{equation}
where $a,e,I$ are the usual orbital elements of the asteroid, and $a_{\rm J}$ is the semi-major axis of Jupiter. 
Asteroids in the Main belt have usually $T_{\rm J}>3$ (Kres\'ak \cite{kres79}).

Note that $T_{\rm J}$ is a constant of motion in the frame of the restricted three body problem 
where the reference plane is that of Jupiter, therefore the inclination must be referred 
to that of the Jovian planet. 



By definition ACOs have orbits with moderate to high inclinations and eccentricities, and have 
low relative speed encounters with Jupiter, {\bf thus it is an unstable population, just like 
comets from the Jupiter family, JFCs, that have dynamical lifetimes of $\sim10^5$ yr (see, for instance,
Alvarez-Candal \& Roig~\cite{alvc04})}. The fact that we can
observe them indicates that the population must be replenished from one or several sources.
The most probable are: {\bf asteroids from the Main belt, which I consider those between 1.8 and 3.2 AU, 
the populations beyond the outer part of the Main belt: 
Cybele ($\sim3.4$ AU), Hilda ($\sim3.9$ AU), Trojan asteroids ($\sim5.2$ AU), 
and JFCs. These last four populations will be collectively called ``primitive'' objects for simplicity.
Here I use the word ``primitive'' with a wide meaning: Those minor bodies whose interiors has not
been widely thermally processed, likely to have a surface (or sub-surface) content of water-ice, and
perhaps undergone water alteration (see chapters 5, 8, and 9 from de Pater and 
Lissauer~\cite{depa07}).} 

The present Research Note is an extension of a work by Licandro et al. (\cite{lica08},
{\bf hereafter L08). {\bf They found that the spectra of ACOs from the visible up to the
near-infrared are alike those of Hilda or Trojan asteroids, but could not set stringent constrains 
on the possible dynamical source of ACOs.}
Nevertheless, they proposed that ACOs with $T_{\rm J}\sim3$ are composed of a mix of {\it bona fide}}
asteroids and primitive object, while going to $T_{\rm J}<2.9$ more primitive objects are found.
I extend their work using a larger database provided by the data from the $4^{\rm th}$ release
{\bf of the Sloan Digital Sky Survey - Moving Objects Catalog (MOC) aiming at a better comparison with 
other primitive populations. In rest of the article I use Sloan photometry as short of
SDSS-MOC photometry.}
In the next section I discuss how to define the sample of ACOs, while in Sec. 3 are 
presented the results, that will be discussed in Sec. 4.

\section{Sloan digital sky survey: Defining the sample}


The Sloan digital sky survey is a large photometric survey mainly developed for stellar and 
extragalactic astronomy. It consists on a set of five magnitudes: $m_u',\ m_g',\ m_r',\ m_i'$, 
and $m_z'$. As a by-product of the reduction pipeline, candidates to moving objects are flagged 
(Ivezi\'c et al. \cite{ivez01}) and are included in the MOC. 
It is then possible to associate these data to real asteroids (Juri\'c et al. \cite{juri02}).
There are more than 100,000 identified asteroids in the MOC, representing almost 50 times more 
objects that the spectroscopic database, slightly over 2000 spectra combining the SMASS 
(Bus \& Binzel \cite{busb02}) and S$^3$OS$^2$ (Lazzaro et al. \cite{lazz04}) databases.

To define the sample of ACOs I selected from the 4$^{\rm th}$ MOC release all observations 
linked to objects with $T_{\rm J} \le3.02$, discarding Centaurs, Trojan, Hilda and Cybele asteroids 
(as in Alvarez-Candal \& Licandro \cite{alvc06} and L08). After this first selection criterion 
a total of 666 observations remained. {\bf The upper cut-off in Tisserand parameter 
is set as that of the JFC 2P/Encke.}

The second step of the selection was based on the relative reflectance computed from
the Sloan magnitudes. They are defined on a set of five filters: {\em u', g', r', i', z'},
centered at 0.354, 0.477, 0.623, 0.763, and 0.913 $\mu$m, respectively. The reflectance 
are computed using 
\begin{equation}
F_j=10^{-0.4[(m_j-m_r')-(m_j-m_r')_{\odot}]},
\end{equation}
where $m_j$ and 
$m_r'$ are the magnitudes in the $j$ and $r'$ filters, respectively, while $\odot$ 
represents the Solar colors (from Ivezi\'c et al. \cite{ivez01}). Note that the reflectance
is normalized to unity at the central wavelength of the $r'$ filter and that they 
represent a very low resolution spectrum.
{\bf The normalization was chosen to make the database compatible
to previous works (such as Roig et al.~\cite{roig08}).}

Using the computed values of reflectance, I eliminated all objects with relative 
errors larger than 10 \%, in {\em g, r, i}, or {\em z}. Anomalous values of flux were 
also discarded, i.e., $F_g'> 1.3$, $F_i' > 1.5$, $F_z'> 1.7$, or $F_g' < 0.6$ (see Roig \& 
Gil-Hutton \cite{roig06}). {\bf The sample was then reduced to 302 measurements, 
including some objects with more that one observation.}

In the final step, I inspected by eye all remaining reflectance, eliminating 
those objects with behavior similar to that of S- or V-type asteroids ({\bf 73 objects, 
generically called ``with bands''}).
I also eliminated those objects with unrealistic reflectance and that survived the 
previous step. 
{\bf The sample was reduced to 111 observations of 94 objects.}

Finally, I computed the spectral slope $S'$ by means of a linear fit to the fluxes $F_{j}$, 
taking into account the errors in each flux, but excluding from the fit $F_{u'}$. 
{\bf The spectral slope is a measure of the redness of the spectrum, the larger the value
the redder the spectrum.} 
{\bf For 
those objects with more than one observation, $S'$ was computed as the weighed mean of each
individual value of $S'$}.
The final sample includes 94 objects with spectral reflectance that resemble those of 
featureless spectra (such as B-, C-, X-, or D-type asteroids).

\section{Results}

I compared the list of 94 ACOs with the spectroscopic sample presented in L08.
There are two objects in common: (6144) 1994 EQ$_{3}$ and (19748) 2000 BD$_{5}$, the 
comparison of the spectral slopes is given in Table \ref{tab01}.
\begin{table}
\begin{minipage}[t]{\columnwidth}
   \caption{ACOs with Sloan and spectroscopic data}
   \label{tab01} 
   \centering                         
   \renewcommand{\footnoterule}{}
   \begin{tabular}{l c c }    
      \hline\hline               
      Object & $S'$ (\% (0.1 $\mu$m)$^{-1}$)  & $S'$ (\% (0.1 $\mu$m)$^{-1}$) \\ 
             & Spec.\footnote{Licandro et al. (\cite{lica08})} & This work  \\ 
      \hline                
       (6144) 1994 EQ$_{3}$ & 7.45 & 9.81 \\
      (19748) 2000 BD$_{5}$ & 5.16 & 1.01 \\
      \hline                               
   \end{tabular}
   \end{minipage}
\end{table}
Considering that error bars of the slopes are typically
about 2 \% (0.1 $\mu$m)$^{-1}$, the values in Table \ref{tab01} are consistent.

\begin{figure}
   \centering
   \includegraphics[width=9cm]{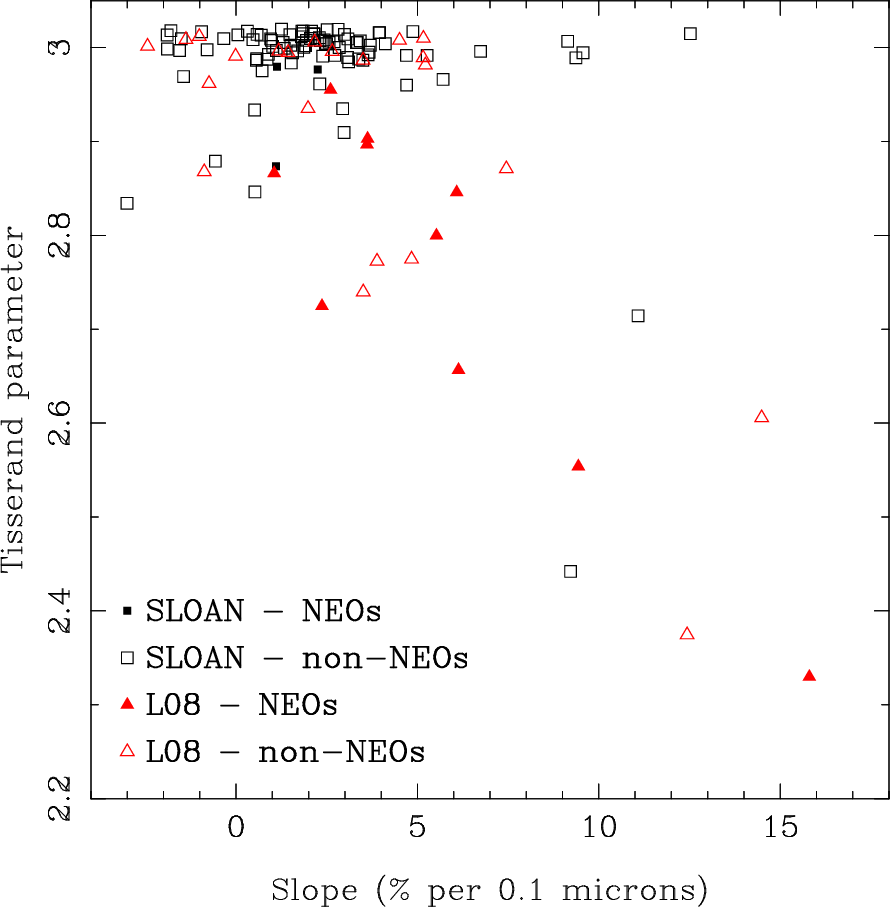}
   \caption{Spectral slope {\it vs.} $T_{\rm J}$. Squares represent ACOs from Sloan while
triangles represent ACOs from L08. 
{\bf Filled symbols indicate objects in near Earth orbits.} In the plot are only represented
objects classified as featureless, see text.}
   \label{fig01}
\end{figure}

In L08 it was found a relationship between $T_{\rm J}$ and $S'$ among the ACOs: Objects with low values of 
$T_{\rm J}$ tend to have the reddest slopes. The spectroscopic database presented in their work 
included 32 featureless ACOs, while now I have a sample that is 3 times larger. Figure 
\ref{fig01} shows the Sloan data and the spectroscopic data in the $S'$-$T_{\rm J}$ space. Note that, 
in order to compare the values of $S'$ of both dataset, I re-computed the slopes of L08 
by re-normalizing them to 0.623 $\mu$m, central wavelength of the {\em r'} filter.

The Sloan slopes follow the same pattern detected with the spectroscopic data, 
confirming therefore the results presented in L08. Nevertheless, among the 
Sloan data there are a few objects with red colors ($\sim10$ \% (0.1 $\mu$m)$^{-1}$) and 
$T_{\rm J}\sim3$, which were not seen in L08. Also, among the new data there are no objects 
as red as the reddest in L08. I do not find any object with $T_{\rm J}<2.8$ and neutral-blue 
slopes. {\bf I tested the apparent correlation seen using the Spearman rank-order correlation
test (Press et al. \cite{pres92}). The test relies on no parameter or {\it a priori} hypothesis.
It computes the correlation coefficient $r$ and its reliability by means of $P_r$. The result
gives $r=-0.27$, $P_r=99.8$ \%, i.e., the anti-correlation seen is statistically reliable
over 3-sigma.}

\begin{figure}[h!]
   \centering
   \includegraphics[width=9.0cm]{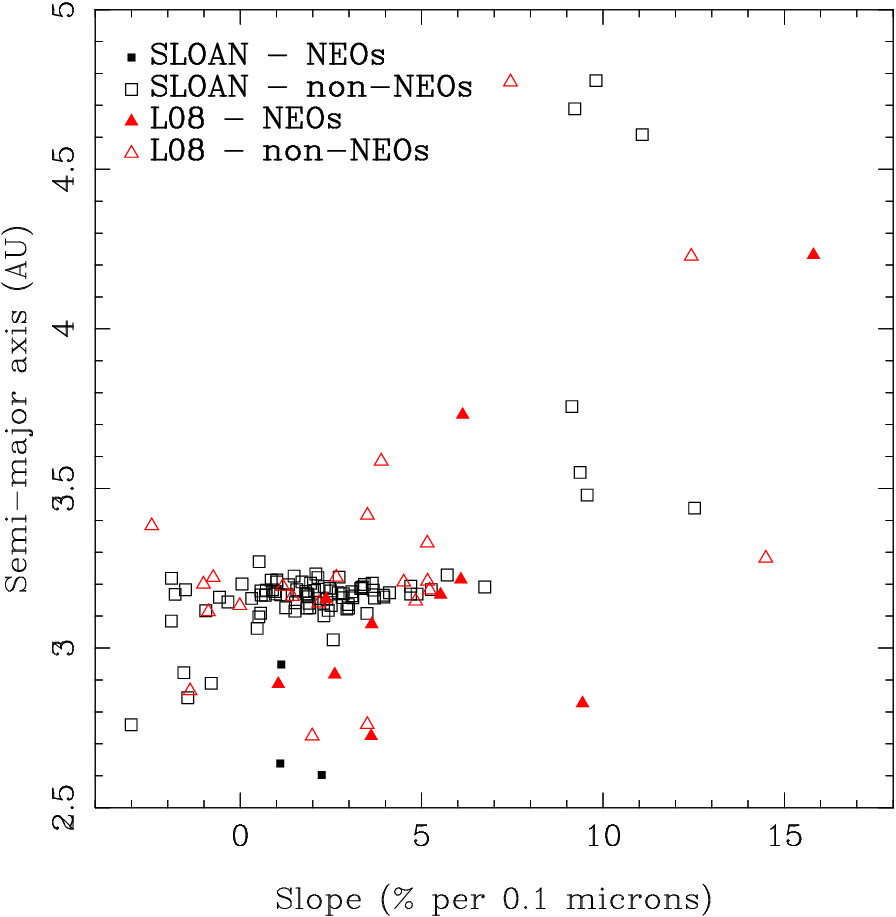}
   \caption{Spectral slope {\it vs.} semi-major axis. The symbol convention is the same as in
Fig. \ref{fig01}.} 
   \label{fig01a}
\end{figure}

\begin{figure}[h!]
   \centering
   \includegraphics[width=9.0cm]{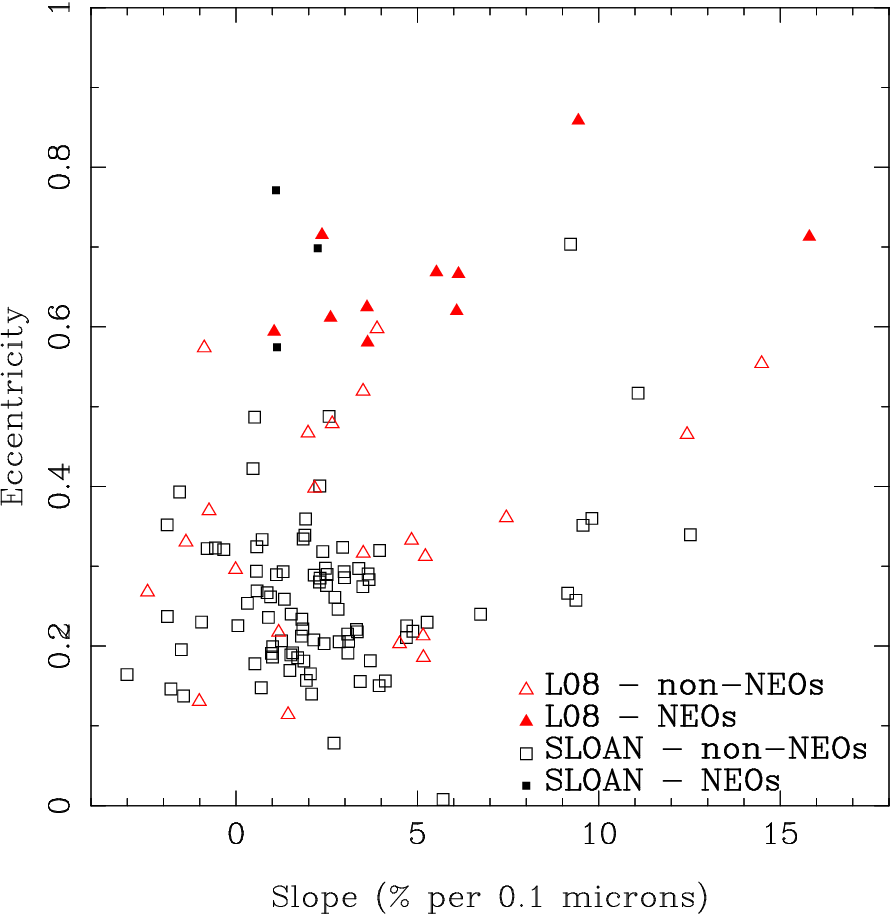}
   \caption{Spectral slope {\it vs.} eccentricity. The symbol convention is the same as in
Fig. \ref{fig01}.} 
   \label{fig01b}
\end{figure}

{\bf I also explored other possible correlations using orbital parameters (Figs. \ref{fig01a} and \ref{fig01b}).
The most significant one is found between spectral slope and eccentricity $r=0.18$ with a reliability
over 2-sigma, which is likely related to the strong correlation found above through the Tisserand parameter 
(Eq. \ref{eq01}).}

\begin{figure}[h!]
   \centering
   \includegraphics[width=9cm]{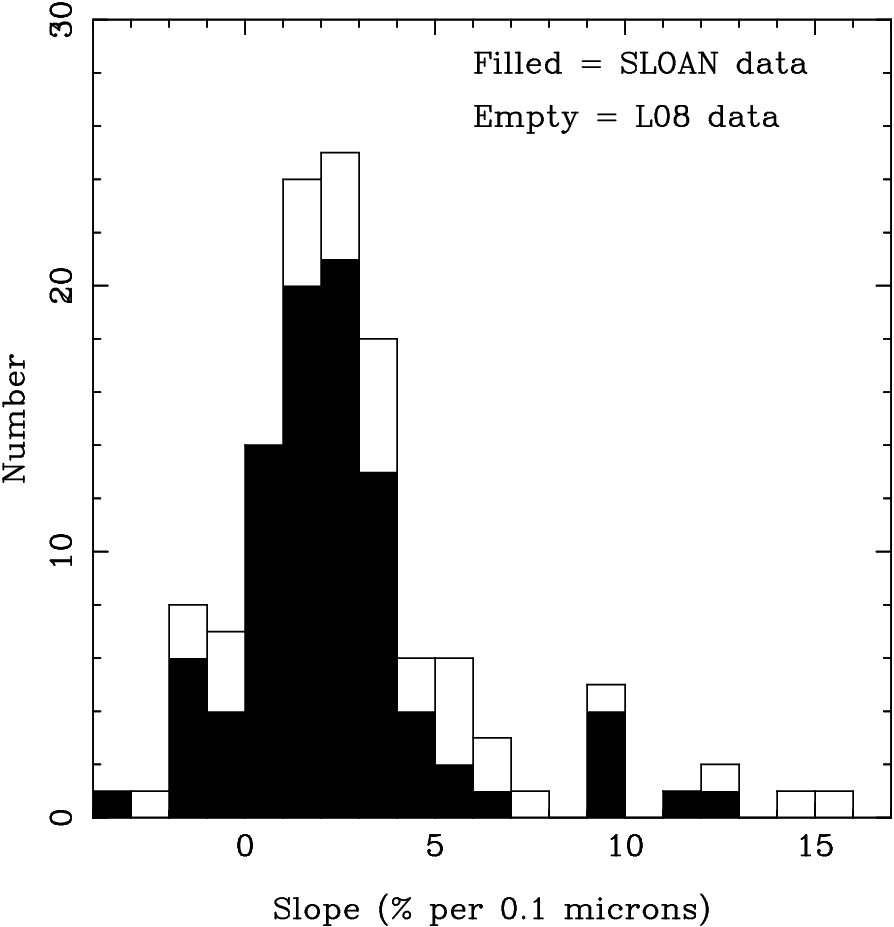}
   \caption{Distribution of spectral slopes for the ACOs. 
The chosen bin is 2 \% (0.1 $\mu$m)$^{-1}$.}
   \label{fig02a}
\end{figure}

A comparison of the distribution of spectral slopes, spectroscopic and Sloan,
is seen in Fig. \ref{fig02a}.
The distributions span about the same range of slopes, with two objects in the 
spectroscopic database redder than any ACO from the Sloan database.

\begin{figure}
   \centering
   \includegraphics[width=9cm]{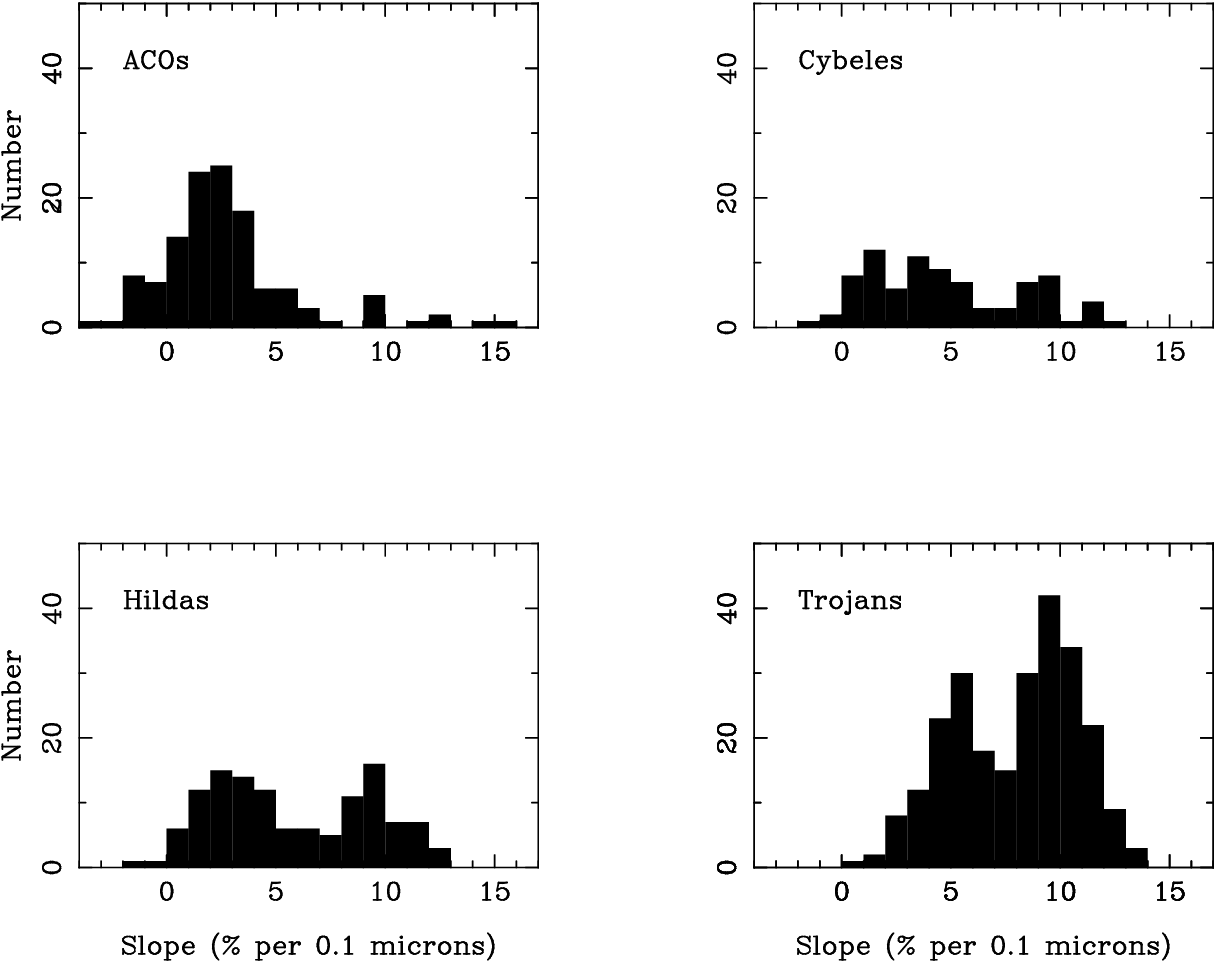}
   \caption{Distribution of spectral slopes for the four different populations. 
{\it Top left:} ACOs, this work plus those from L08 - 
{\it Top right:} Cybele asteroids from Gil-Hutton \& Licandro \cite{gilh09} - 
{\it Bottom left:} Hilda asteroids from Gil-Hutton \& Brunini \cite{gilh08} - 
{\it Bottom right:} Trojan asteroids from Roig et al. \cite{roig08}.
The chosen bin is 2 \% (0.1 $\mu$m)$^{-1}$.}
   \label{fig02}
\end{figure}

Three works study primitive populations of the Solar System using Sloan photometry: 
Gil-Hutton \& Brunini (\cite{gilh08}), Roig et al. (\cite{roig08}), and Gil-Hutton \& Licandro
(\cite{gilh09}) studying the Hilda, Trojan, and Cybele asteroids, respectively.
I compare the distribution of spectral slopes of the three samples in Fig. \ref{fig02}.

The distributions are different, spanning more or less the same range of slopes. 
The ACOs have a majority of objects with slopes
$0-5$ \% (0.1  $\mu$m)$^{-1}$, which, according to Fig. \ref{fig01} have $T_{\rm J}\sim3$.
Nevertheless, there is a tail of objects with $S'\geq10$, which coincides with the 
reddest objects from the other populations.

{\bf I ran a Kolmogorov-Smirnov test (Press et al. \cite{pres92}) comparing the ACOs slope 
distribution with each one of the other three distributions presented in Fig. \ref{fig02} 
testing the null hypothesis that they have been extracted from the same parent distribution.
The results came out negative in each case.}

\section{Discussion}
\begin{figure}
   \centering
   \includegraphics[width=9cm]{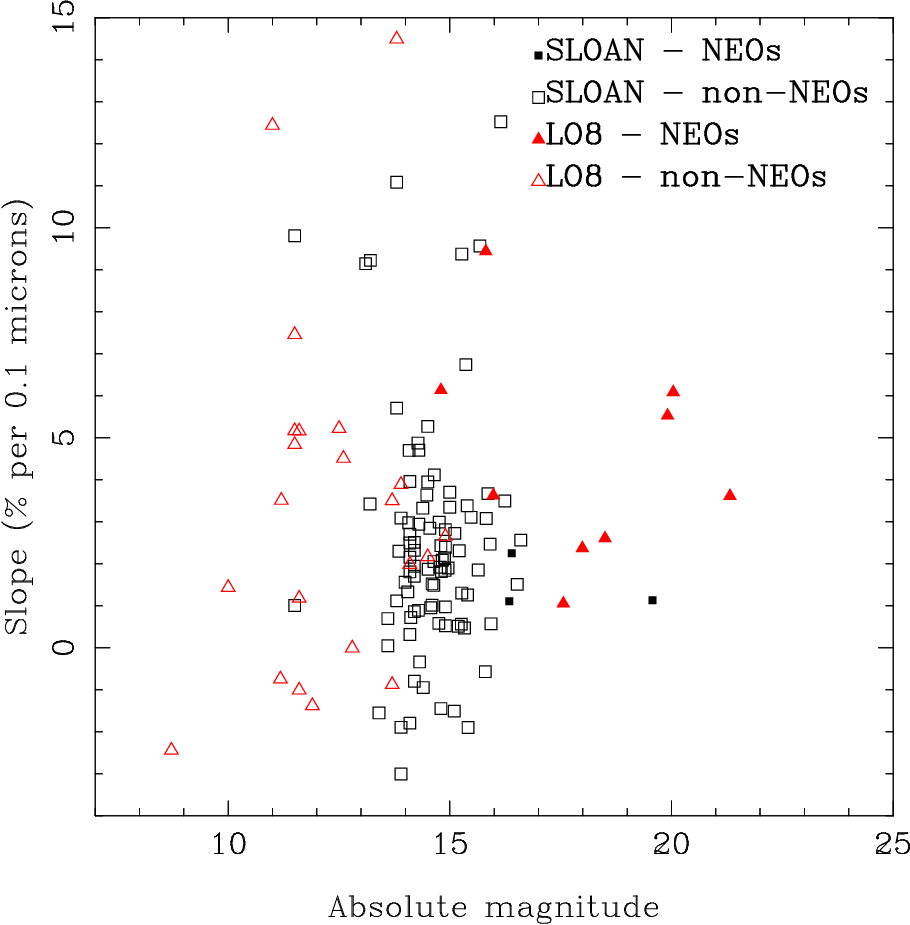}
   \caption{ Absolute magnitude {\it vs.} spectral slope for the ACOs' samples. The
symbols code is the same as those in Fig. \ref{fig01}.}
   \label{fig03}
\end{figure}

Using the Sloan digital sky survey data I increased by a factor three the 
number of ACOs with physical observations pushing the limiting magnitude of detection
by one (see Fig. \ref{fig03}). Note that this applies only to ACOs in non NEO orbits. 
{\bf Figure \ref{fig03} should be regarded as illustrative of gain in number by using the Sloan database.
Nevertheless, it should be kept in mind that the sample is affected by at least two selection effects:
The first one against near Earth objects: By observational strategy an object must be 
observed with all five filters to be flagged as moving object candidate,
which is not the case for near Earth orbits which have high speed and cross the field-of-view faster than
the rate at which the telescope switches filters. The second one is against farther, darker, 
smaller object.}

The slope distribution of ACOs could reflect the distributions from the contributing
populations. Dynamical studies indicate that Trojan and Hilda asteroids could contribute 
to the population of asteroids in cometary orbits (Levison et al. \cite{levi97}, Di Sisto 
et al. \cite{disi05}). As the ACO distribution of slopes does not resemble those of Cybele,
Hilda, or Trojan asteroids (Fig. \ref{fig02}), they are not the only contributors, probably not
even the main ones, therefore the Main belt or JFCs are probably important too.
{\bf Figure \ref{fig01a} shows that most of the ACOs with neutral or slightly red spectral
slopes (C or X taxonomical classes) locate just below 3.2 AU, the outer limit of the Main belt.
These objects are also located mostly around $T_J\sim3$ and probably have their origin within the 
Main belt.}

As a by-product of the distributions of slopes, we can see that there is an increase
of the average value of $S'$ with increasing distances from the Sun:
{\bf 2.8 \% (0.1 $\mu$m)$^{-1}$, 4.8 \% (0.1 $\mu$m)$^{-1}$, 5.8 \% (0.1 $\mu$m)$^{-1}$, 
and 8.13 \% (0.1 $\mu$m)$^{-1}$, for ACOs, Cybele, Hilda, and Trojan asteroids, respectively.}

\begin{figure}
   \centering
   \includegraphics[width=9cm]{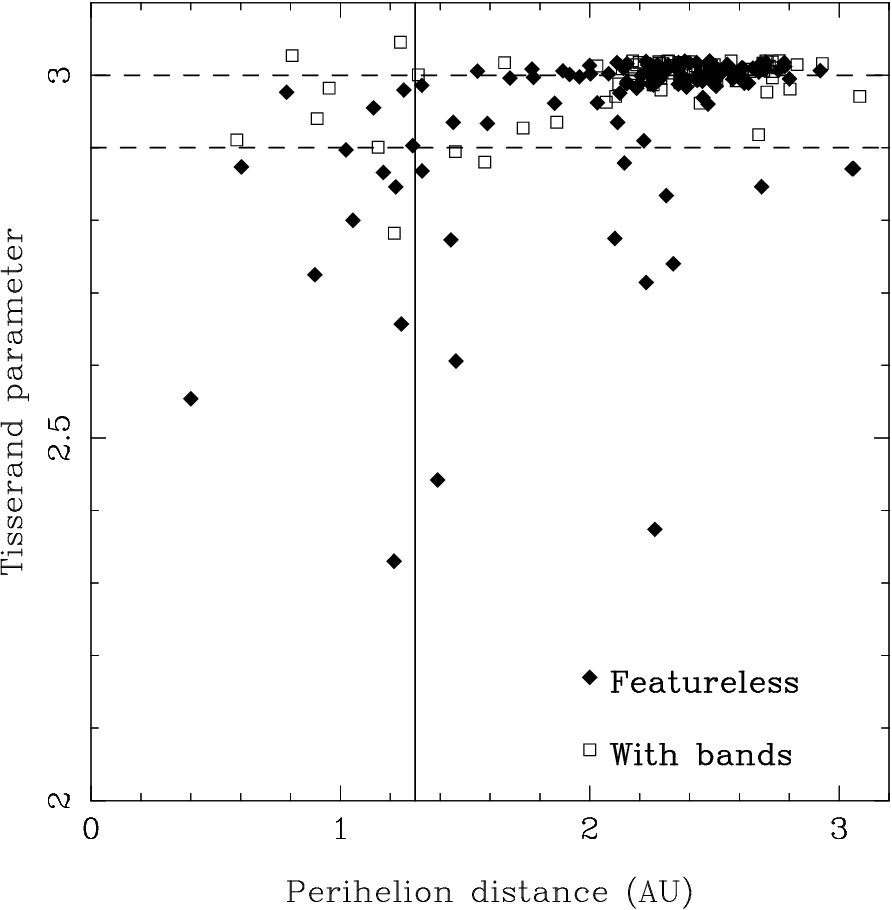}
   \caption{Perihelion distance {\it vs.} Tisserand parameter. The open squares indicate ACOs with bands. 
Filled rhombus indicate featureless ACOs. Continuous line separate the near earth objects 
(NEOs, $q<1.3$ AU) from the non-NEOs. Dashed horizontal lines indicate $T_{\rm J}=2.9$ and $T_{\rm J}=3.0$. All ACOs
from L08 and this work are considered together.} 
   \label{fig04}
\end{figure}

{\bf Figure \ref{fig04} reproduces Fig. 3 from L08. It shows all ACOs in the sample:
featureless and with bands, as described above. The figure shows that 
most of the ACOs with bands have $T_{\rm J}>2.9$, with very few exceptions.}
This confirms the result presented in L08 with a smaller database. {\bf The population at 
$T_{\rm J}\sim3$ is probably a mix of asteroids from the inner and outer regions
of the Main belt and a few primitive objects, while at lower values of $T_{\rm J}$ are probably 
objects that were injected into those orbits from primitive populations. I remind the reader
that the convention used in this work calls primitive populations these outside the Main belt,
i.e., farther than 3.2 AU, as well as Jupiter family comets.}

Considering together the Sloan and L08 samples, and remembering that there are two objects
in common, there are observations for a total of 204 ACOs, 124 of them have featureless spectra.
In the sub-population of NEOs there are 21 objects, 14 of which are featureless (67 \%,
in good agreement with DeMeo \& Binzel \cite{deme08}'s estimative). A similar
fraction, 60 \% of the objects, are featureless in the non-NEO population.
Note that, when considering only objects with $T_{\rm J}<2.9$, the fractions change radically,
so does the number of observed ACOs. Globally, 24 out of 27 objects are featureless
(89 \%). The fraction of featureless ACOs is more or less constant for, either, the NEO 
(10 out of 11) or non-NEO (14 out of 16) populations.

It is known that the near-Earth region is a mixture of different populations of minor bodies: 
Main belt and primitive populations (Bottke et al. \cite{bott02}). 
When I consider objects that have $T_{\rm J}<2.9$ the results
indicate a lower contribution from asteroids from the Main belt
where objects with bands predominate (see Moth\'e-Diniz et al. \cite{moth03}).

To the best of my knowledge, no dynamical study has been carried out about the possible 
origins of the ACOs in non-NEO orbits. 
Nevertheless, the great mix found for $T_{\rm J}>2.9$ 
indicates a mixture of different populations, perhaps similar to what is seen for NEOs.
{\bf The Main belt population of asteroids has typical values of $T_{\rm J}>3.0$, thus 
it remains to be understood how those objects had their orbits excited to $2.9<T_{\rm J}<3.0$.
One possibility are resonant perturbations.} Once more, as with the NEOs, objects
which have $T_{\rm J}<2.9$ had, for the most part, their sources in the primitive populations.

Other open question is what is the link between the ACOs dynamical evolution and the 
physical-chemical evolution of their surfaces. 
Figure \ref{fig04} hints that most of the objects with features in their spectra, coming 
from the Main belt of asteroids, are not able to dynamically evolve to values of $T_{\rm J}<2.9$, therefore
reinforcing the possibility that the ACOs with $T_{\rm J}<2.9$ have their source from the primitive
populations.

\begin{acknowledgements}
I thank F. Roig and R. Gil-Hutton for kindly providing their databases of Trojan,  
Hilda, and Cybele asteroids. The author also acknowledges support from the Marie
Curie Actions of the European Commission (FP7-COFUND). I also thank F. DeMeo whose
comments and critics improved this manuscript.
\end{acknowledgements}

\end{document}